\documentclass[12pt]{article}
\usepackage{graphicx}
\begin{document} 

\title{Monte Carlo simulation of survival for minority languages}

\author{Christian Schulze and Dietrich Stauffer\\
Institute for Theoretical Physics, Cologne University\\D-50923 K\"oln, Euroland}

\maketitle
\centerline{e-mail: stauffer@thp.uni-koeln.de}

\bigskip
\small{
Abstract: Our earlier language model is modified to allow for the survival of 
a minority language without higher status, just because of the pride of its
speakers in their linguistic identity. An appendix studies the roughness of the 
interface for linguistic regions when one language conquers the whole territory.
}


\bigskip

\section{Introduction}

Canada is a multicultural country with two official languages:
English (majority) and French (minority). How is it possible
that the minority language does not die out even though it 
has no higher prestige (status) there than the majority language 
English? We want to simulate this effect.

Abrams and Strogatz \cite{abrams} introduced competition between
languages in a model with only two languages, and there the 
minority language can survive only if it has a higher status 
(prestige) than the majority language. With some modification 
also coexistence is possible \cite{argentina} but still a status 
advantage is needed. Other aspects of language competition were
recently reviewed by us \cite{chachacha}. The present work uses 
our previous multi-language model \cite{cise} to explain a stable
survival of one minority language without an overall status
advantage. 

\section{Model}

The model \cite{cise} is a variant and generalisation of our 
earlier bit-string model \cite{chachacha} and uses Potts
variables $q = 1,2,...,Q$ instead of merely bits. Each language
or grammar is characterised by $F$ features, and each of these
features is an integer between 1 and $Q$. Thus we have $Q^F$
possible languages or grammars. 

People sit on $L \times L$ square lattices, one person and one
language per lattice site. Initially each person speaks a
randomly selected language. Then, at each iteration, three
processes mutation, transfer and flight happen with probabilities
$p,q,r$:

Each of the $F$ features is mutated with probability $p$. If 
such a mutation happens, then with probability $1-q$ a random
integer between 1 and $Q$ is selected as the new value for this 
feature, while with probability $q$
one of the four lattice neighbours is selected randomly and its
value for this feature is transferred to become the value of 
this feature for the mutated language. If the language is spoken
by a fraction $x$ of the whole population, then with probability
$r(1-x)^2$ the simulated person switches its language to that 
of a randomly selected lattice neighbour. (In our earlier 
version \cite{cise} this flight away from a small language was
made to the language of a person randomly selected from the whole
lattice instead of from the four neighbours; in that case we
do not get coherent language regions.) Usually, $r = 0.9$.

As in earlier versions, this model gives a sharp phase transition
where the equilibrium fraction of people speaking the largest 
language jumps from a small value to nearly unity, that means the
population moves from fragmentation (into many languages) to 
dominance (of one language). 

We now assume that the speakers of one particular language, which
may be identified with French in case of Canada, start to
defend their language as soon as the largest lšlanguage (English)
has attracted more than half of the population. From then on 
the flight from French to other languages no longer takes place, 
that means $r = 0$ for all French speakers. The other languages
(native population, other European immigrants, ...) continue
as usual towards near-extinction.

\begin{figure}[hbt]
\begin{center}
\includegraphics[angle=-90,scale=0.3]{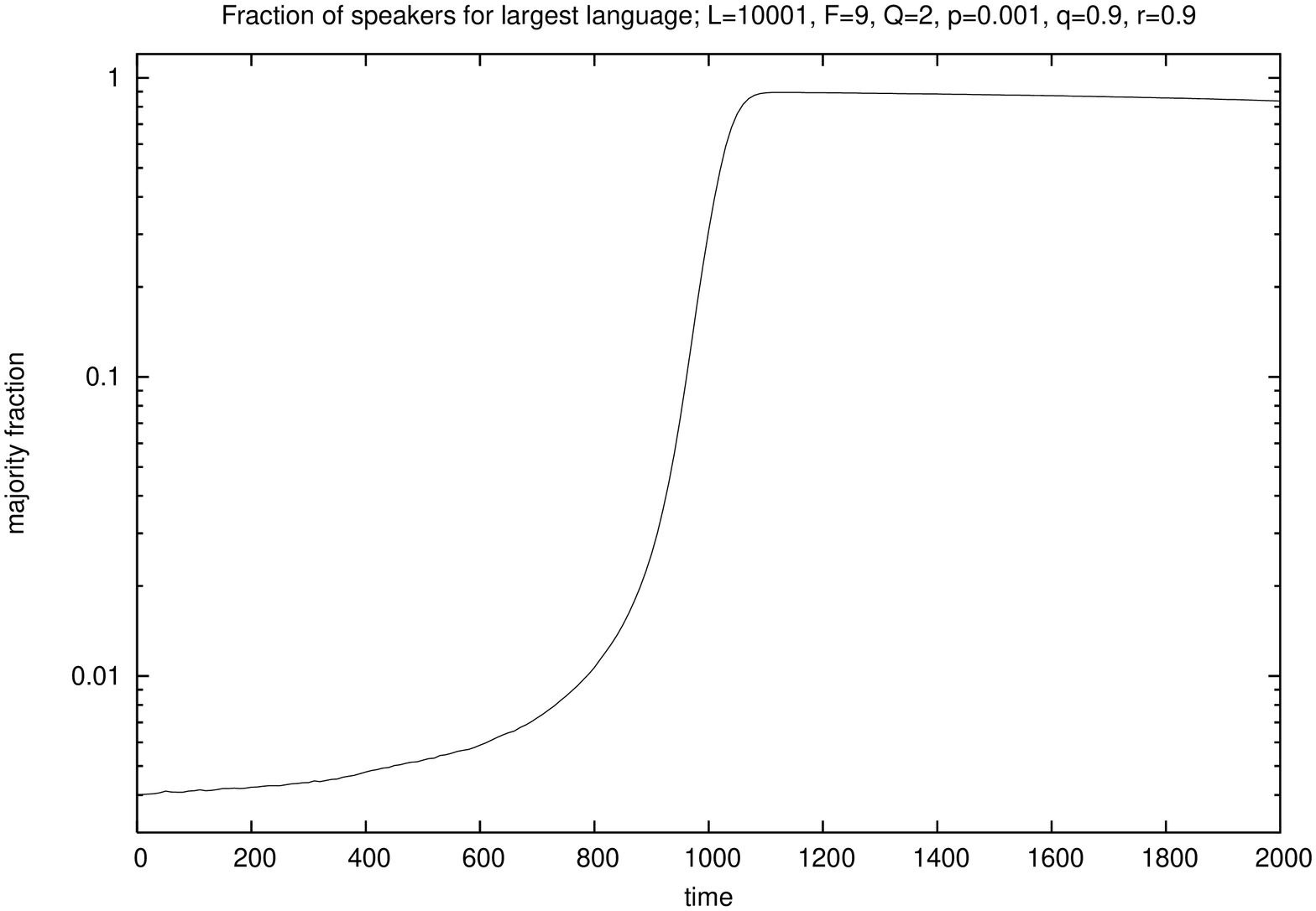}
\includegraphics[angle=-90,scale=0.3]{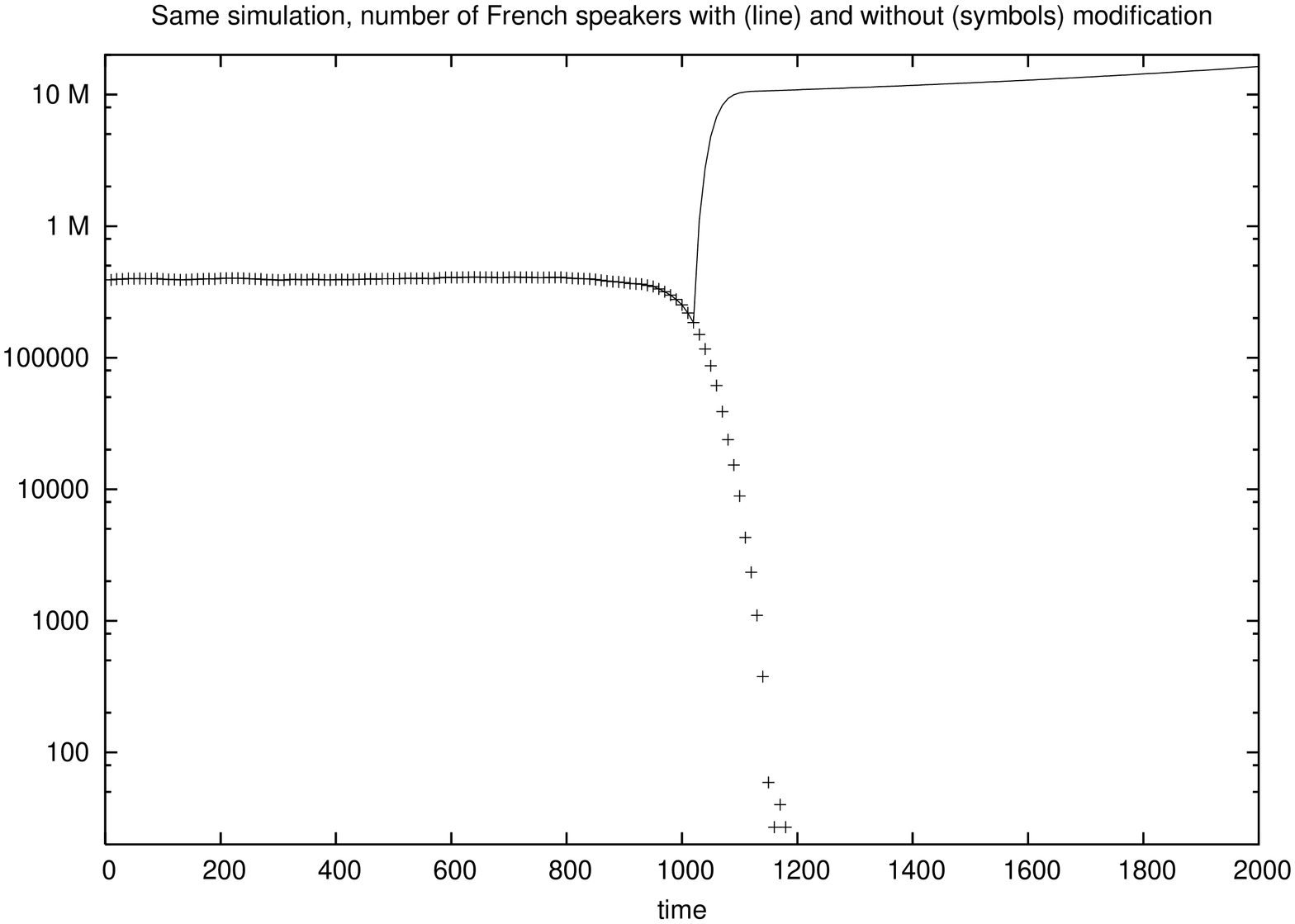}
\end{center}
\caption{Time variation of ``English'' (top) and ``French'' (bottom) with
100 million people, with and without the modification regarding flight from 
French.
}
\end{figure}

\begin{figure}[hbt]
\begin{center}
\includegraphics[angle=-90,scale=0.3]{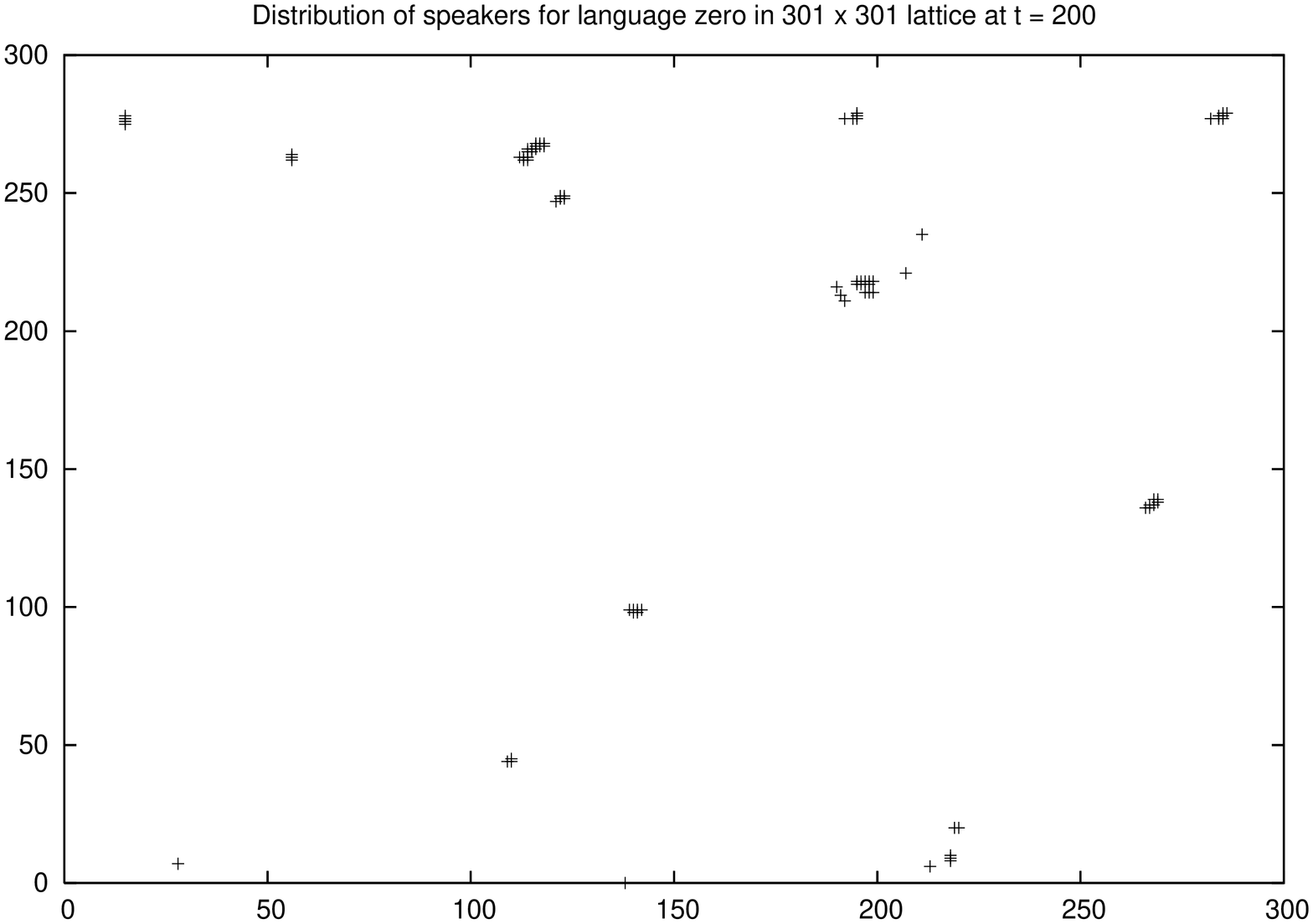}
\includegraphics[angle=-90,scale=0.3]{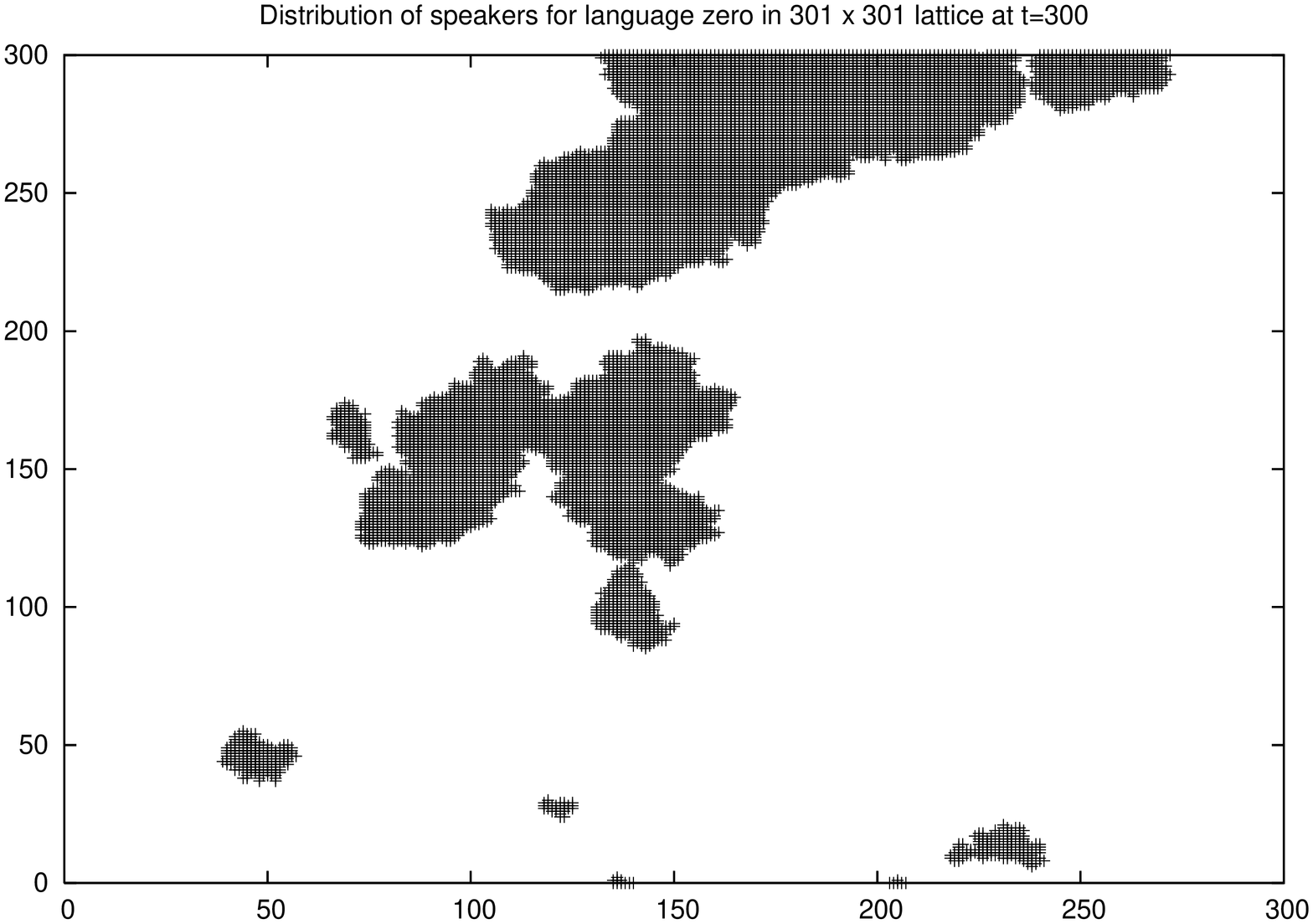}
\end{center}
\caption{Clusters of French-speaking people before (top) and after (bottom) 
English became the dominating language and French speakers no longer left 
their language. As in Fig.1: $F = 8,\; Q = 2,\; p = 0.001, \; q = 0.9, \; r=0.9.$
}
\end{figure}

\begin{figure}[hbt]
\begin{center}
\includegraphics[angle=-90,scale=0.5]{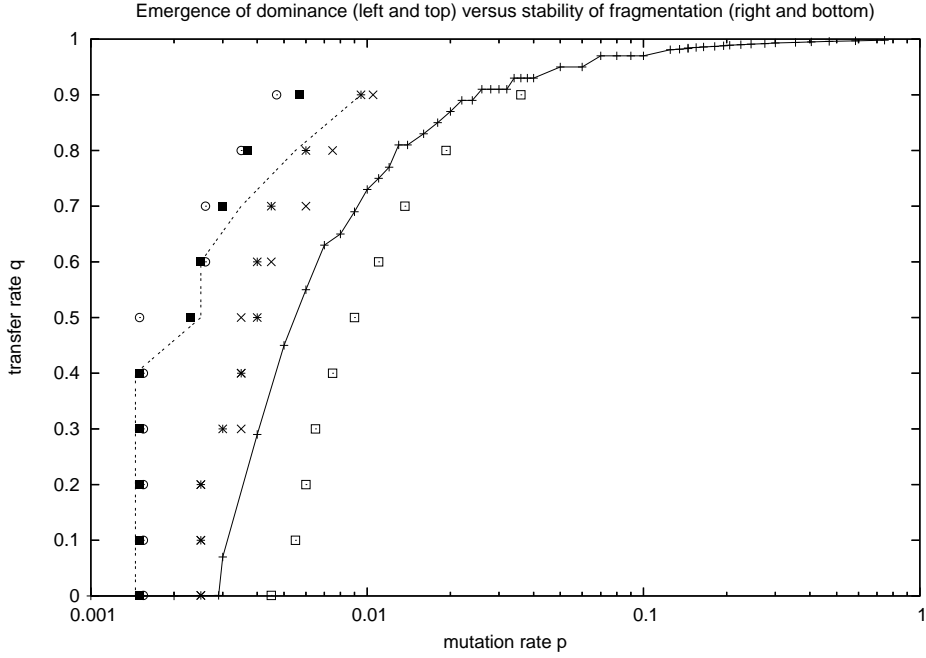}
\end{center}
\caption{Phase diagram for $L = 201, \; t = 10,000, \; r = 0.9$. To the left 
we get dominance to the right we stay at fragmentation. $(F,Q)$ = (8,2; empty 
square), (8,3; x), (8,5; *),  (16,2; full square), (16,3; empty circles) from
right to left. The lines have $F = 8, \; Q = 2$ for $L = 1001$ (one 
sample) instead of 201 (four samples), with the usual $r=0.9$ for the long line
and $r=0.5$ for the short line. 
}
\end{figure}

\begin{figure}[hbt]
\begin{center}
\includegraphics[angle=-90,scale=0.5]{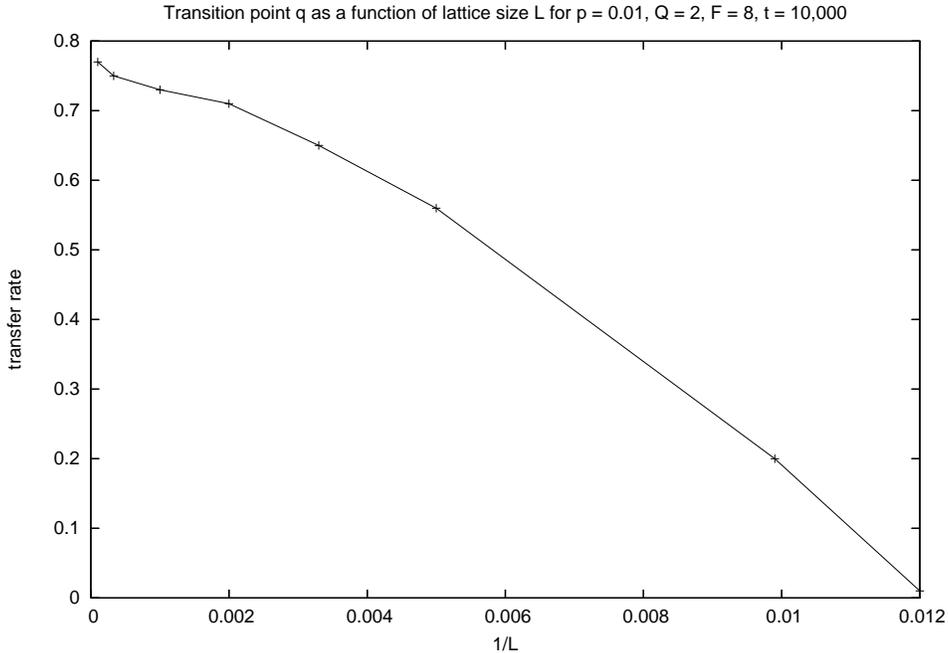}
\end{center}
\caption{Strong size effects for transition point $q$ with $83 \le L \le 10001$. 
}
\end{figure}

\begin{figure}[hbt]
\begin{center}
\includegraphics[angle=-90,scale=0.5]{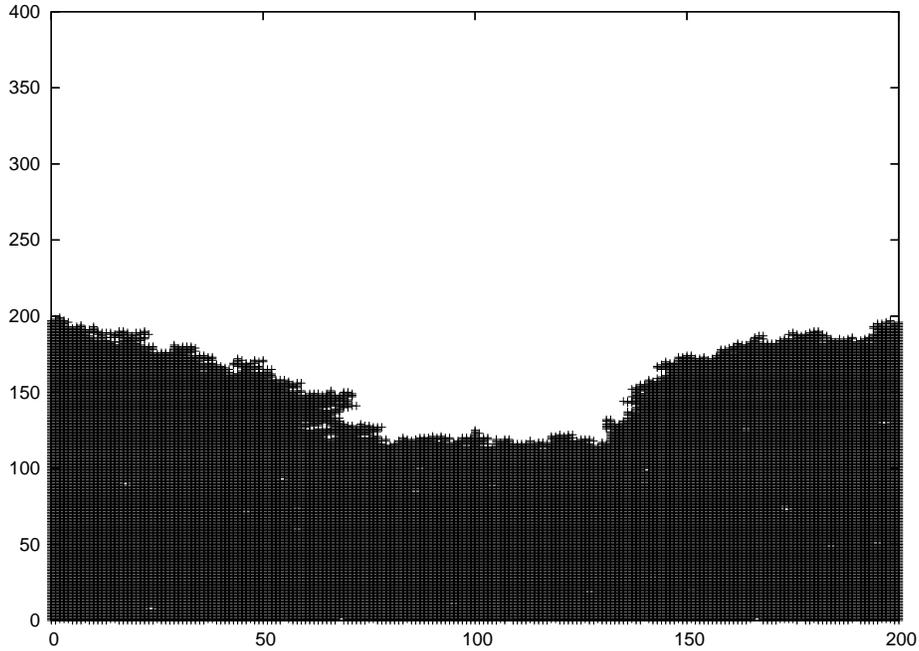}
\end{center}
\caption{Growth of the region where the dominating language is spoken, at 
intermediate times. At the end nearly everybody speaks that language}
\end{figure}

\begin{figure}[hbt]
\begin{center}
\includegraphics[angle=-90,scale=0.26]{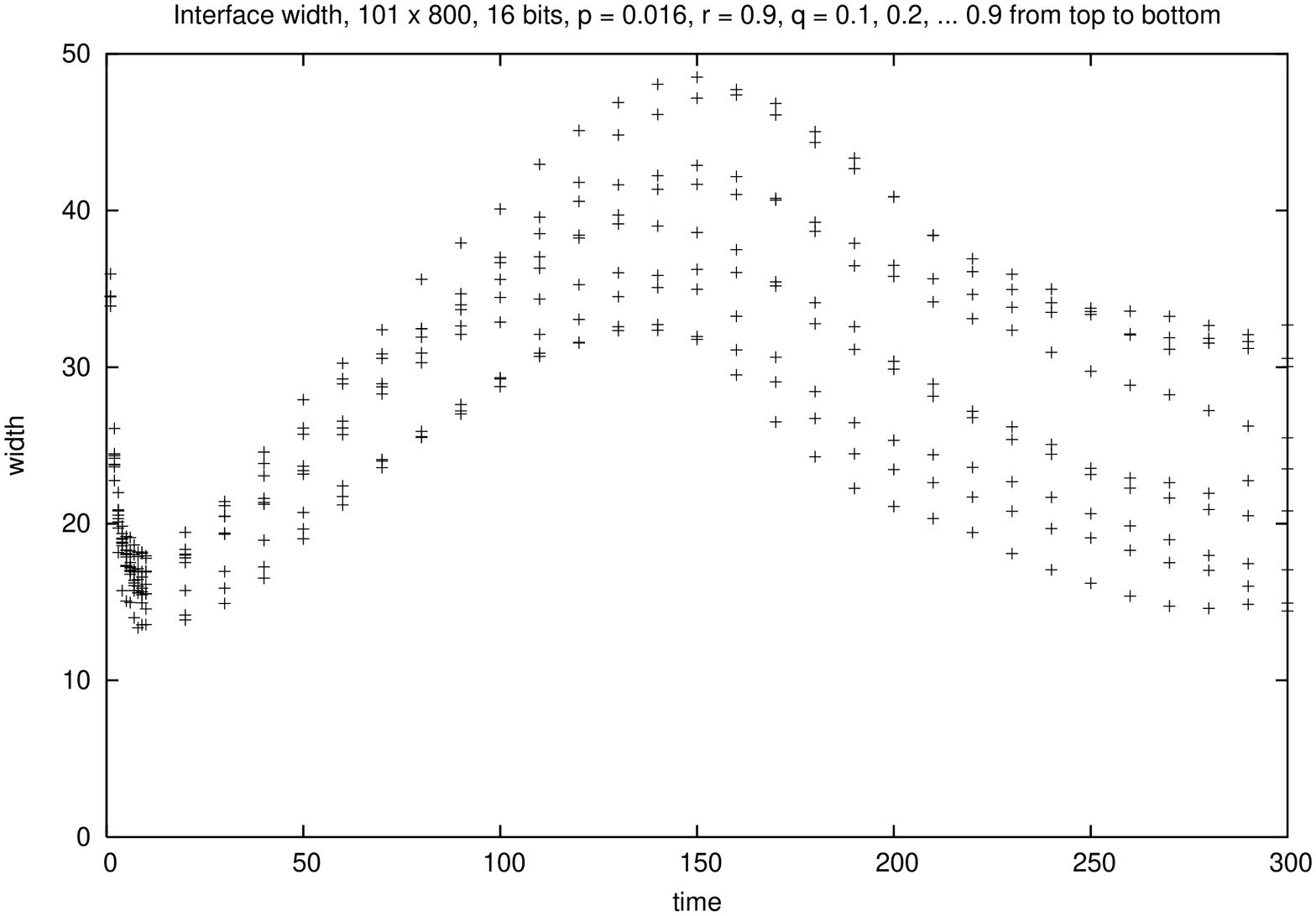}
\includegraphics[angle=-90,scale=0.26]{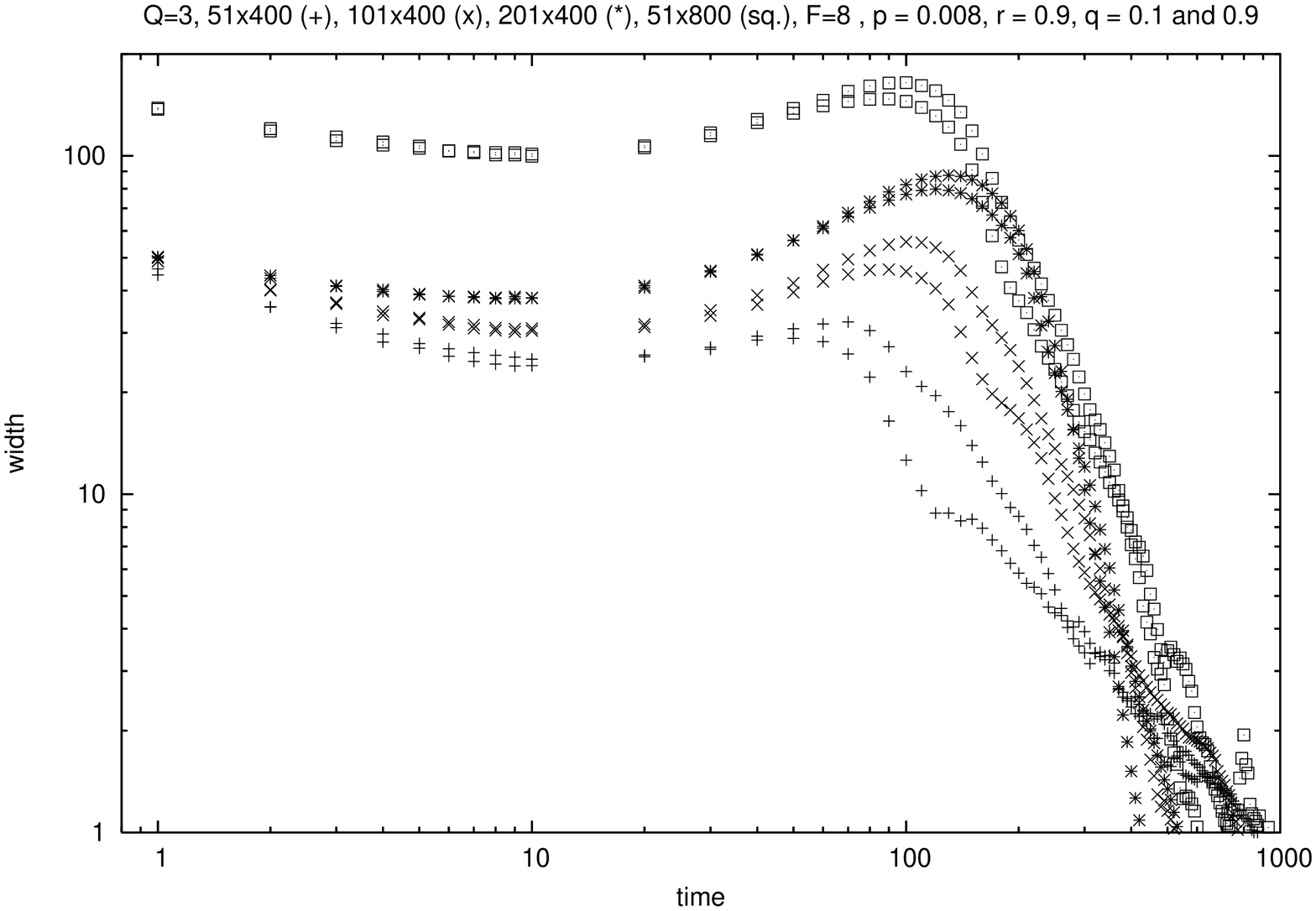}
\includegraphics[angle=-90,scale=0.26]{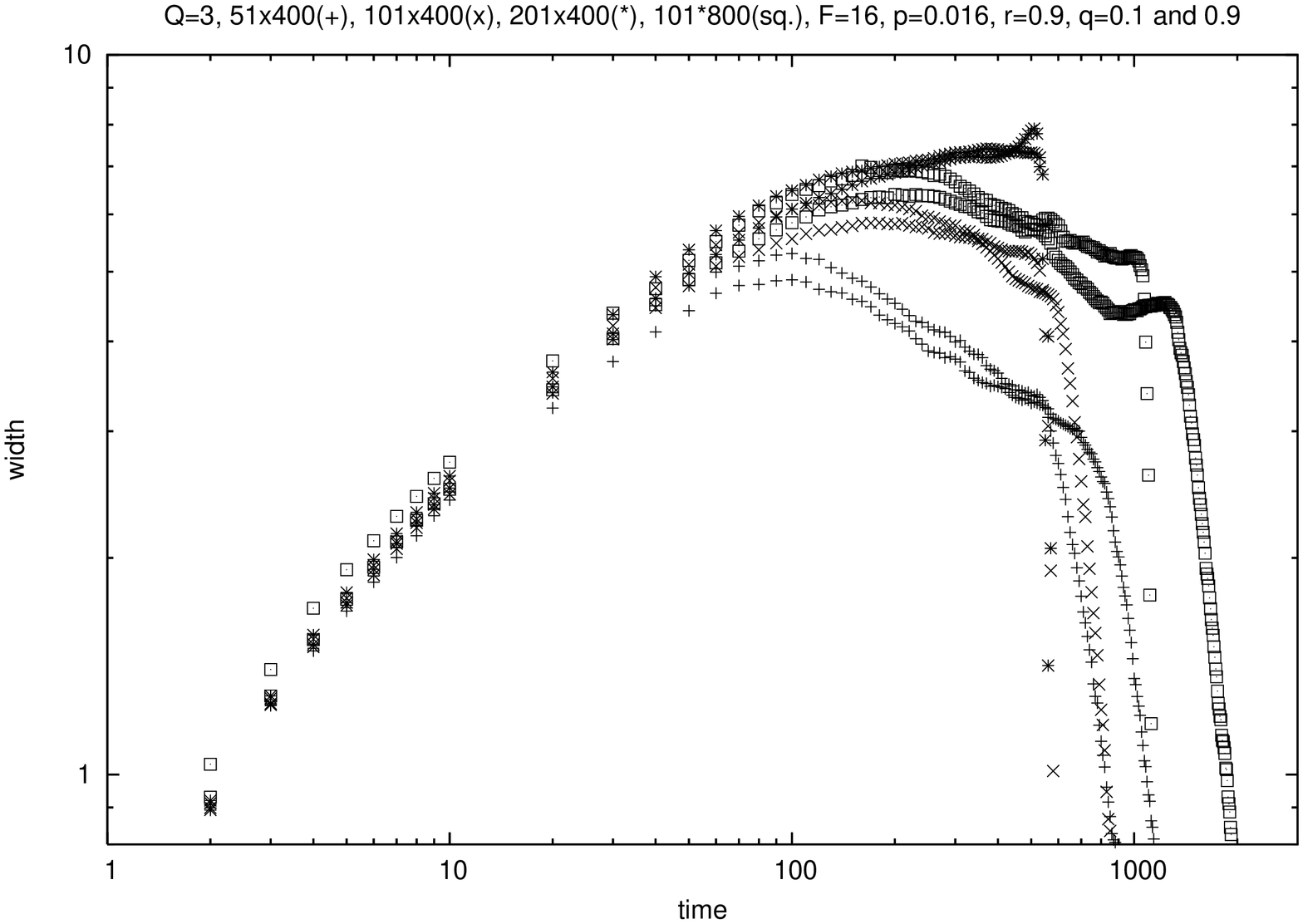}
\includegraphics[angle=-90,scale=0.26]{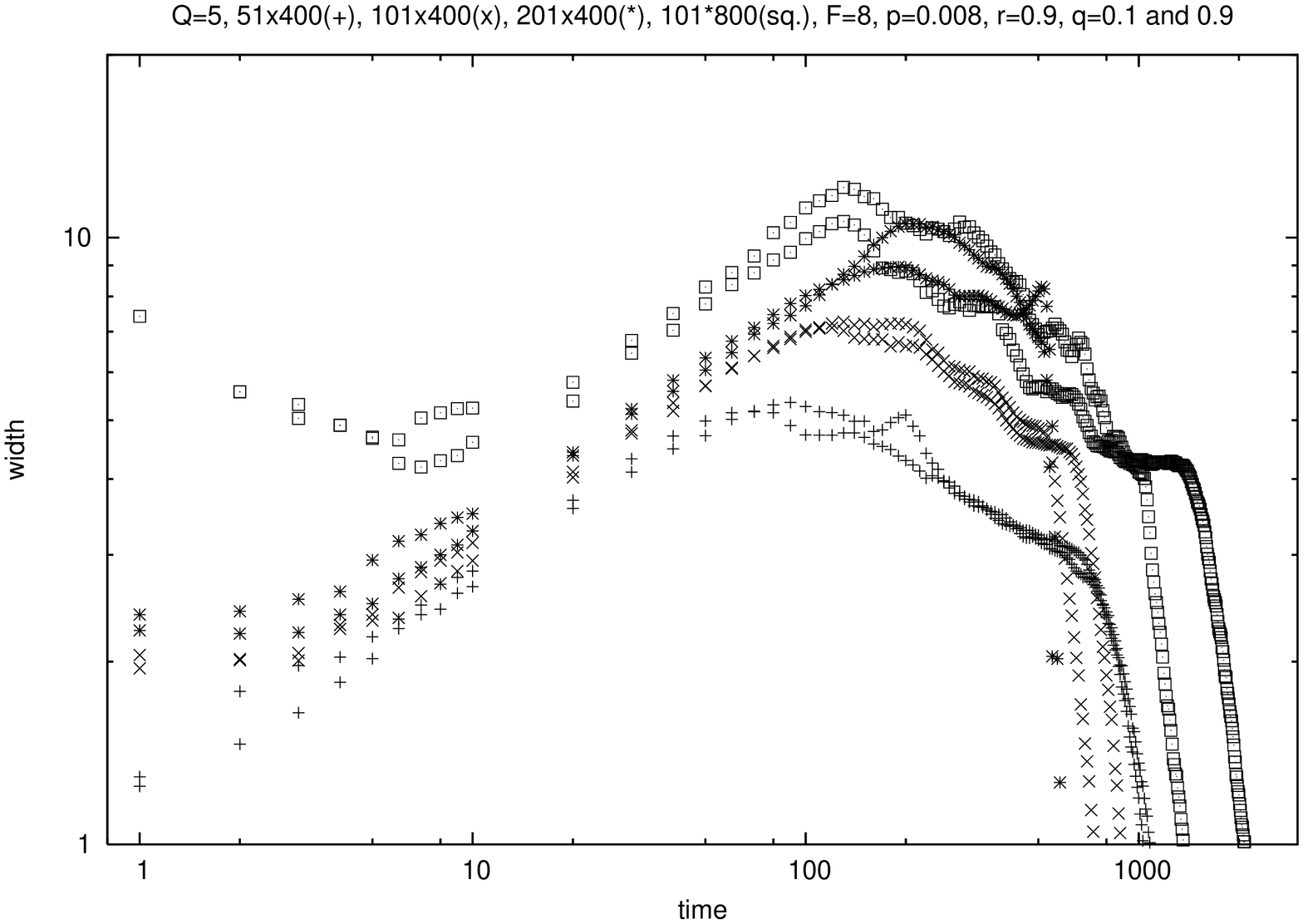}
\end{center}
\caption{Width versus time for $Q = 2$ (part a), $Q = 3$ (parts b,c) and 
$Q = 5$ part d), with $F = 8$ (parts b,d) and 16 (parts a,c). For $Q = 2,
\; F = 8$ (not shown) the width decays rapidly to zero.
}
\end{figure}

\section{Results}

Fig.1 shows the growth of the majority language, and Fig.1b
the variation of the special language (i.e. French); initially
both are spoken by about 0.4 percent of the population. We see
that after slightly more than 1000 iterations, when English 
reaches the 50 percent threshold, the French language recovers 
from its previous losses and attracts even more people than 
initially, though it remains less widespread than English (except 
perhaps for very long times).  

Fig.2 shows for a smaller lattice with $10^5$ instead of $10^8$
sites that French speaking people form geographically
connected clusters, both before and after English surpasses the
50 percent threshold at t = 244. The French clusters before 
this threshold time are much smaller than afterwards.   

These simulations were made for small mutation rates $p=0.001$,
where dominance always emerges independent of $q$. For larger
$p$, dominance is possible only for large $q$, and Fig.3 shows 
the phase diagram: In the left part (small $p$ or large $q$)
fragmentation switches over to dominance while in the right part
(large $p$ or small $q$) fragmentation stays during the 
observation time of $10^4$ iterations. (Errors are about 0.02 in 
$q$ at fixed $p$ and 0.001 in $p$ at fixed $q$.)
We see little qualitative dependence on these parameters, but the more 
possible languages we simulate (the higher $F$ or $Q$ is), the more difficult is
the emergence of dominance from the initial fragmentation.
Fig.4 shows unusually large finite-size effects: Only for $L > 82$ 
a transition to dominance could be seen at $p = 0.01$; 
nevertheless a finite transition point $q_c \sim 0.76$ seems 
plausible for infinitely large lattices. (As a function of 
observation time $t$, at $L = 2, \; F=8, \; Q=2, \; p = 0.01, \; r = 0.9$ 
the threshold $q_c$ diminished from 0.65 and 0.56 to 0.52 and 0.48 for
$t = 10^3, \ 10^4, \ 10^5, \ 10^6$.)

One might envisage a situation when the use of French dies out
even after the flight away from French is stopped. This happened
for our parameters if French had died out already before the
flight from French was stopped. Such a situation is nearly 
unavoidable for large $F$, large $Q$ and small $L$, when there 
are much more possible languages $Q^F$ than people $L^2$. 

\section{Conclusion}

We modified our previous multi-language model by allowing flight 
away from small languages only to languages spoken by a lattice
neighbour, and by switching off this flight for one particular
language (French) once the dominating language (English) 
surpassed a threshold of half of the population. As a result, 
the French language did not only survive (if it did not become
extinct before) but even could attract more speakers than at
the initial random distribution of the population among all 
possible languages. 

\section{Appendix}

The irregular clusters of language domains in Fig.2 are not suited to study
surface roughening, familiar to physicists since decades. To study the 
structure of the interface between French and the other languages we thus
started with the bottom five percent of the lattice speaking only French, and 
omitted the rule that the French stop switching to English once more than half 
of the population speaks English. Because of our initialisation with five 
percent, French is the dominating language anyhow and the no-switch condition 
is never fulfilled. The simulation now corresponds more to the boundary
between a conquering language and native different languages, like in Quebec
three centuries ago. Fig.5 shown an intermediate example of the growth of 
French. At the end the border merges with the upper lattice line and nearly 
everybody speaks French.

We now define a width $W$ of the interface region between black and white in 
Fig.5 by determining for each column $k$ the highest black place $i_k$ where 
French is spoken. Then our width is the mean square line number $i$:
$$  W^2 = \sum_k (i_k - <i>)^2/H; \quad <i> ,\ = \sum_k i_k/H$$
where $H$ (up to 800) is the height of the $L \times H$ lattice. Fig.6 shows 
for various parameters the variation of $W$ with $L, \; H, \; q$; that with $q$ 
is quite small. For our standard case $Q = 2, \; F=8$ we found a linear increase
of $W$ with time until the finite height hinders further growth, Fig.6a; for 
the other parameters $Q=3$ and 5, and $F = 16$, the log-log plots Fig.6b-d are 
less clear. Perhaps the First Nations in Quebec were not interested in the 
exponents of the Kardar-Parisi-Zhang equation.

\end{document}